\documentclass[twocolumn,prl,aps]{revtex4}

\usepackage{graphicx}
\usepackage{amsmath}
\usepackage[french]{babel}
\usepackage[latin1]{inputenc}

\begin{document}

\title{The gauge non-invariance of Classical Electromagnetism}
\author{Germain Rousseaux}
\affiliation{Institut Non-Lin\'{e}aire de Nice, Sophia-Antipolis.\\
UMR 6618 CNRS. \\ 1361, route des Lucioles. 06560 Valbonne, France. \\
(Germain.Rousseaux@inln.cnrs.fr) }

\begin{abstract}
{\it Physical theories of fundamental significance tend to be gauge theories. These  are theories in which the physical system being dealt with is described by more variables than there are physically independent degree of freedom. The physically meaningful degrees of freedom then reemerge as being those invariant under a transformation connecting the variables (gauge transformation). Thus, one introduces extra variables to make the description more transparent and brings in at the same time a gauge symmetry to extract the physically relevant content. It is a remarkable occurrence that the road to progress has invariably been towards enlarging the number of variables and introducing a more powerful symmetry rather than conversely aiming at reducing the number of variables and eliminating the symmetry} \cite{Henneaux}. We claim that the potentials of Classical Electromagnetism are not
indetermined with respect to the so-called gauge transformations. Indeed, these transformations raise paradoxes that imply their rejection. Nevertheless, the potentials are still indetermined up to a constant.

\end{abstract}

\maketitle

\section{Introduction}

In Classical electromagnetism, the electric field ${\bf E}$ and the
magnetic field ${\bf B}$ are related to the scalar $V$ and vector ${\bf A}$ potentials by the following definitions \cite{Jackson} :
\begin{equation}
{\bf E}=-\frac {\partial {\bf A}} {\partial t}-\nabla V
\quad and \quad
{\bf B}=\nabla \times {\bf A}
\end{equation}

One century ago, H.A. Lorentz noticed that the electromagnetic
field remains invariant ($\bf{E'}=\bf{E}$ and $\bf{B'}=\bf{B}$)
under the so-called gauge transformations \cite{Okun}:

\begin{equation}
{\bf A'}= {\bf A} + \nabla f
\quad and \quad
V'=V - \frac {\partial f} {\partial t}
\end{equation}
where $f(x,t)$ is the gauge function.

Hence, this indeterminacy is believed to be an essential
symmetry of Classical Electromagnetism \cite{Okun}. Moreover, it
is often related to the assertion that the potentials are not
measurable quantities contrary to the fields. Hence, one must
specify what is called a gauge condition, that is a supplementary equation which
is injected in the Maxwell equations expressed in function of the electromagnetic
potentials in order to supress this indeterminacy. It is common to say that these gauge conditions are
mathematical conveniences that lead to the same determination of the
electromagnetic field. In this context, the choice of a specific gauge condition is
motivated from the easiness in calculations compared to another one. In a certain manner, although their mathematical expressions are different, it is supposed that they are equivalent as the fields are invariant with respect to the gauge transformations. Furthermore, no physical meaning is ascribed to the gauge conditions as the potentials are assumed not to have one...

Despite these assertions which are shared by a large majority of physicists, a definition for the potentials  dating back to Maxwell was recalled recently and which resolves, according to our point of view, the question of indeterminacy by giving them a physical interpretation. Moreover, we showed  that the Coulomb and Lorenz gauge conditions were, in fact, not equivalent because they must be interpreted as physical constraints that is electromagnetic continuity equations \cite{Guyon,Rousseaux}. In addition, we were able to demonstrate that the Coulomb gauge condition is the galilean approximation of the
Lorenz gauge condition within the magnetic limit of L\'evy-Leblond \& Le Bellac \cite{Levy,ARQS,RL}. 
So, to "make a gauge choice" that is choosing a gauge condition is, as a consequence of our findings, 
not related to the fact of fixing a special couple of potentials. Gauge conditions are completely uncorrelated to the supposed indeterminacy of the potentials. Hence, we proposed to rename "gauge condition" by "constraint" \cite{Guyon,Rousseaux,RL}.

In this article, we would like to reexamine the common belief concerning the assumed indeterminacy of the potentials with the assumption that the "constraint" do not fix the value of the potentials. Indeed, we will show that gauge transformations introduce paradoxes which imply their rejection. This point of view was expressed already by the school of De Broglie : by the master himself \cite{Broglie} or by his followers like Costa De Beauregard \cite{Costa}...

\section{The case of a stationnary electric field}

Imagine a one dimensional {\bf stationary}
problem defined by the following potentials :

\begin{equation}
{\bf A= 0}
\quad and \quad
V(x)=-Ex
\end{equation}

One finds easily :
\begin{equation}
{\bf B= 0}
\quad and \quad
{\bf E}= E {\bf e_x}
\end{equation}
The electric and magnetic fields are constants in time.

Now, we can perform a gauge transformation with this particular
gauge function :

\begin{equation}
f(x,t) = -Ext
\end{equation}

The new potentials are :

\begin{equation}
{\bf A'}=-{\bf E}t
\quad and \quad
V' = V- \frac{\partial f }{\partial t } = 0
\end{equation}

Of course, the electric field is unchanged but is the underlying
physics expressed by the potentials the same ? We believe that an
electric field can be created by two very different physical
processes that is time variation of a vector potential (like in
induction phenomena) or space variation of a scalar potential
(like in the electron gun). We are in front of the first paradox :
how can a physical quantity (here, the vector potential) be a
function of time in a stationary problem ? In the case of a capacitor for example, the static electric field is created by static electric charges on the plates of the capacitor. If one admits that one can describe this static electric field by a time-dependent vector potential, one must admit that the sources (charges or currents) of this electric field are time-dependent which is not experimentally the case.

The unconvinced reader could argue that we proposed a gauge
function which depends explicitly on time in a stationary problem.
As a matter of fact, if $f(x,t)$ does not depend on time, the
scalar potential is invariant $V'=V$ and so is not indetermined
with respect to the gauge transformations.

In a stationary problem, the electric field is expressed as ${\bf
E}=-\nabla V$ whereas the magnetic field is still defined as
previously. In this case, the vector potential could still be
indetermined. So, are two vector potentials differing from a
gradient physically equivalent ?

\section{The case of a vector potential equal to a gradient}

Now, one will apply the so-called Stokes-Helmholtz-Hodge decomposition to the vector potential :
\begin{equation}
{\bf A}={\bf A_{longitudinal}} + {\bf A_{transverse}}
\end{equation}
with :
\begin{equation}
{\bf A}= \nabla g + \nabla \times {\bf R}
\end{equation}
where $g$ is a scalar and $\bf R$ a vector. The decomposition is unique up to the additive gradient of a harmonic function with the following properties \cite{Helmholtz}:
\begin{equation}
\nabla .{\bf A_{harmonic}} = {\bf 0} \quad \nabla \times {\bf A_{harmonic}} = {\bf 0}
\end{equation}

If we use gauge transformations, we can notice that only the
longitudinal (and/or harmonic) part of the vector potential and the scalar potential
are affected by these transformations. The transverse part remains
unchanged. Moreover, the magnetic field depends only on the
transverse part. So, if there is indeterminacy, it must imply
indeterminacy of the longitudinal (and/or harmonic) part. As a consequence, the
longitudinal (and/or harmonic) part cannot have a physical meaning if it is
indetermined with respect to the gauge transformations.

Usually, the vector potential is equal to its transverse part in
most of the problems of Classical Electromagnetism. For example, the vector potential for a magnet is expressed by :
\begin{equation}
{\bf A}= {\bf A_{transverse}} = \frac{\mu _0}{4\pi} \nabla \times( \frac{{\bf m}}{r})
\end{equation}
where ${\bf m}$ is the strengh of the poles (the so-called magnetic mass or moment).

In this case, we observe a magnetic field by definition. And, if the vector potential varies in
time, it creates an electric field again by definition : the time
integral of the electric field could be considered as a direct measure of
the vector potential without the presence of static charges that is of a scalar potential. If not, one can use the superposition theorem to evaluate first the part of the electric field associated to the static charge (that is the scalar potential) and then the part associated to the current (that is the vector potential) if and only if the separation is possible...

At this stage, the question is to know whether a vector potential only equal to a gradient can have a physical effect when the electromagnetic field is null. 

Outside a solenoid, the vector potential is precisely equal to a gradient as expressed by the
following formula in cylindrical coordinates $(r,\theta)$ \cite{Jackson} :
\begin{equation}
{\bf A} =\frac {\Phi}{2 \pi r} {\bf e_\theta} =\frac {\Phi \nabla \theta} {2 \pi}
\end{equation}
where $\Phi$ is the flux of magnetic field inside the solenoid or
the circulation of the vector potential outside the solenoid. More precisely, its curl and divergence are null so the vector potential outside a solenoid is of a harmonic-type according to the indeterminacy of the Stokes-Helmholtz-Hodge decomposition \cite{Helmholtz}. Moreover, we point out forcefully that the mathematical indeterminacy due to the gauge transformations is discarded by the boundary conditions which give a physical determination to the vector potential outside a solenoid : the vector potential vanishes far from its current sources.

The external region of a solenoid is one of the numerous experimental configurations to observe
 the well-known Aharonov-Bohm effect \cite{Feynman}. So, we are in front of the second paradox :
the Aharonov-Bohm effect contradicts the fact that a vector
potential equal only to a gradient could not have a physical effect.

However, one usually argues that the Aharanov-Bohm effect
is a quantum effect and that the potentials can have a meaning in
quantum physics and not in classical physics. Moreover, the
prediction of the Aharonov-Bohm effect pointed out an
unanticipated way in which the vector potential can affect a
measurement on a charged particle in a region of zero field, but
the predicted phase is the result of a path integral which insures
that the result is gauge independent : it is not the result of a
local (gauge-dependent) value of the vector potential.

Why add in the path integral a (mathematical) gradient to the
vector potential which is already a (physical) gradient ?

A vector potential equal only to a gradient should not have a physical effect as imply
by gauge transformations because we could cancel the longitudinal (and/or harmonic)
part by adding the gradient of the appropriate gauge function. 

In the solenoid example, we can take as a gauge function
$f(r,\theta,t) =-\Phi \theta/2 \pi$ and still there is
experimentally an effect despite the fact that all the potentials
and so the fields cancel outside the solenoid.

However, one can found in the litterature some theoretical arguments against this gauge transformation according to the fact that the existence of the solenoid implies that the the space is not simply-connected. Yet, it is true that in a multiply connected region, the function $\alpha$ which characterises the longitudinal (and/or harmonic) part of the vector potential in the general case becomes multivalued but the longitudinal (and/or harmonic) part of the vector potential (the only one which is different from zero outside the solenoid) is not multivalued as one take the gradient of $\alpha$. 

Another argument is to remember that the gauge transformations were introduced by Lorentz without any constraint on the connectness of the space.

Having the Aharonov-Bohm effect in mind, we can recall now
 a very simple experiment which cannot be explained with Maxwell equations expressed in
function of the electromagnetic field only and which shows the physical
character of a longitudinal (and/or harmonic) vector potential in classical physics.

Let's take again the geometry of the solenoid. If the current
varies with time the magnetic field is still null outside the
solenoid but because the vector potential is not null outside the
solenoid and varies with time, it creates an electric field
outside the solenoid. If we denote the flux of the magnetic field
inside the solenoid (or the circulation of the vector potential
outside the solenoid) $\Phi=LI$ where $L$ is the inductance of the
solenoid and $I$ the current intensity, the electric field is
expressed by :
\begin{equation}
{\bf E}=-\frac {\partial {\bf A}} {\partial t}=-{\frac {L}{2\pi r}} \frac{dI}{dt}{\bf e_\theta}
\end{equation}

If we apply Maxwell equations expressed in function of the fields
with the prescription that the magnetic field is null outside the
solenoid, we only find that the electric field is lamellar outside
the solenoid which is supposed to be infinite ($\nabla \times {\bf E}=0$ because ${\partial
\bf{B}}/ {\partial t}=\bf{0}$ even in this time-dependent problem
because $\bf{B}=\bf{0}$ outside the solenoid)... 

This experiment is carried out very easily. It demonstrates that a
vector potential only equal to a gradient can have a physical
effect in Classical Electromagnetism when it varies in time and
thus creates by definition an electric field. Of course, for a finite solenoid, the leaking magnetic field is not null outside the solenoid. However, it creates a leaking electric field which is negligeable and opposite with respect to the electric field created by the contribution of the vector potential due to the ideal solenoid...

Another example of a physical vector potential which is equal to a gradient appears in
the well-known Meissner effect in supraconductivity and it was discussed nicely by Tonomura \cite{Tonomura}.

\section{The case of a uniform magnetic field}

Another drawback of the gauge transformations can be illustrated
by the following example : one often finds in textbooks that we can
describe a uniform magnetic field ${\bf B} =B {\bf e}_{z}$ by either the so-called
symmetric "gauge" $\bf{A_s}=1/2\bf{B} \times \bf{r}$ or by the
so-called Landau "gauge" \cite{Okun}. This two "gauges" are related
by a gauge transformation :

\begin{equation}
{\bf A}_{1}=\frac{1}{2} {\bf B} \times {\bf r} = \frac{1}{2} [-By,Bx,0]
\end{equation}
becomes either :
\begin{equation}
{\bf A}_{2} =[0,Bx,0]
\quad
or
\quad
{\bf A}_{3}=[-By,0,0]
\end{equation}
with the gauge functions $\pm f = \pm xy/2$.

However, there is no discussion in the litterature of the following 
issue. As a matter of fact, if we consider a solenoid with a current along
$\bf{e_\theta}$, the magnetic field is uniform (along $\bf{e_z}$)
and could be described by the symmetric "gauge" or the Landau
"gauge". Yet, the vector potential in the Landau "gauge"  ${\bf A}_{2}$ is along
$\bf{e_y}$ whereas the vector potential in the symmetric gauge is
along $\bf{e_\theta}$. We advocate that only the symmetric "gauge"
is valid in this case because it does respect the symmetry of the 
currents (${\bf J} = J {\bf e}_{\theta}$) whereas the Landau "gauge" 
does not. Moreover, the symmetric "gauge" (or the Landau "gauge") is not, in fact, a gauge condition
 but a solution describing a uniform magnetic field under the Coulomb constraint
($\nabla .{\bf{A_1}}=\bf{0}$).
In order to understand this last point, one can picture an analogy between Fluid Mechanics and Classical Electromagnetism. Indeed, the solenoid is analogous to a  cylindrical vortex core with vorticity $\bf{w}$ and we know that the velocity inside the core is given by
$\bf{u}=1/2\bf{w} \times \bf{r}$ which is analogous to the
symmetric gauge for an incompressible flow ($\nabla .{\bf{u}}=\bf{0}$). Outside the vortex core, the velocity is given by \cite{GHP} :
\begin{equation}
{\bf u}=\frac {\Gamma \nabla \theta} {2
\pi}=\frac {\Gamma}{2 \pi r} {\bf e_\theta}
\end{equation}
where $\Gamma$ is the flux of vorticity inside the vortex or the
circulation of the velocity outside the vortex. One recovers the
analogue formula for the vector potential outside a solenoid...

Of course, if the problem we are considering does not feature the cylindrical geometry (two horizontal plates with opposite surface currents for example, analogous to a plane Couette flow \cite{GHP}), one of the Landau gauges ${\bf A}_{2}$ or ${\bf A}_{3}$ must be used instead of the symmetric gauge ${\bf A}_{1}$ according to the necessity of respecting the underlying distribution/symmetry of the currents which is at the origin of both the vector potential and the magnetic field. To give a magnetic vector field without specifying its current source is an ill-posed problem which was interpreted so far by attributing an indeterminacy to the vector potential which is wrong.

Now, how can we test experimentally this argument based on symmetry ? If the current of the solenoid varies with time, it will create an electric field which is along $\bf{e_\theta}$ as the vector potential because the electric field is minus the time derivative of the vector potential. If the currents in the horizontal plates change with time, a horizontal electric field will appear for the same reason. The author rejects all the arguments based on the fact that one can define through a gauge transformations a time dependent scalar potential which would explain the oberved electric field. Indeed, a scalar potential is physically defined with respect to charge distributions and not current distributions.

Contrary to the common belief, it is possible to discriminate experimentally between two vector potentials (related by a gauge transformation) creating a uniform magnetic field. We have shown that the symmetry of the current source implies a certain distribution of the vector potential which is at the origin of an electric field when the intensity is time-dependent. Its orientation is dictated by the vector potential
alone which, eventhough is not observable by itself, has observable consequences.

\section{Conclusions}

What is the meaning of gauge transformations ? We believe that it
is only a structural feature (that is linearity) of the
definitions of the potentials from the fields. The potentials of
Classical Electromagnetism do have a physical meaning as recalled
recently \cite{Guyon,Rousseaux,RL,Mead,K} and should be considered as the starting point of Classical Electromagnetism \cite{RL,Mead}. If we defined the fields from the potentials
and not the contrary, the gauge transformations loose their sense
because they imply the paradoxes raised in this article. As a
conclusion, we must reject gauge transformations. Gauge invariance
is preserved but in a weaker sense : the potentials are defined up
to a constant. This constant is equal to zero when the sources are confined to a certain region of space : one assumes that the potentials vanish at infinity far from their sources. If the domain is bounded like in a Faraday cage, the surface potentials are given by the contribution of all the sources outside the region of interrest \cite{Mourier} : one makes the assumption that their knowledge is not important as one measures only differences of potentials according to their definition in function of a	constant of reference \cite{RL}. We recall for example that the vector potential is an electromagnetic impulsion that is a difference of electromagnetic momentum. In mechanics, a momentum is indetermined as it is defined with respect to a reference but the impulsion constructed from this momentum has a definite value so is not indetermined...

\end{document}